\documentclass[conference]{IEEEtran}
\IEEEoverridecommandlockouts

\newcommand\blfootnote[1]{%
  \begingroup
  \renewcommand\thefootnote{}\footnote{#1}%
  \addtocounter{footnote}{-1}%
  \endgroup
}

\usepackage{cite}
\usepackage{adjustbox}
\usepackage{amsmath,amssymb,amsfonts}
\usepackage{graphicx}
\usepackage{textcomp}
\usepackage{multicol}
\usepackage{multirow}
\usepackage{subcaption}
\usepackage{enumitem}
\usepackage{balance}
\usepackage{booktabs}
\usepackage{array}
\usepackage[table,xcdraw]{xcolor}
\newcolumntype{x}[1]{>{\centering\arraybackslash\hspace{0pt}}p{#1}}

\usepackage{pst-all}
\usepackage{pstricks-add}
\usepackage{float}

\usepackage{pgfplots}

\usepackage{tikz}
\usepackage{tikzscale}
\usepackage{transparent}

\usepackage{algorithm}
\usepackage[noend]{algpseudocode}

\usepackage{comment}

\usepackage{blindtext}

\definecolor{purplelight}{HTML}{ead1dc}
\definecolor{purpledark}{HTML}{a64d79}
\definecolor{colorRed}{HTML}{E19794}
\definecolor{colorGreen}{HTML}{b6d7a8}
\definecolor{colorBlue}{HTML}{a4c2f4}

\def\BibTeX{{\rm B\kern-.05em{\sc i\kern-.025em b}\kern-.08em
    T\kern-.1667em\lower.7ex\hbox{E}\kern-.125emX}}
\begin{document}

% Computing Resource Allocation Utilizing Community Detection and Matching Technique of Social Internet-of-Things

\title{Graph Neural Networks-based Clustering for Social Internet of Things\vspace{-0.5cm}}
\author{
\IEEEauthorblockN{Abdullah Khanfor$^1$,
  Amal Nammouchi$^1$,
  Hakim Ghazzai$^1$,
  Ye Yang$^1$,
  Mohammad R. Haider$^2$,
  and Yehia Massoud$^1$\\
\IEEEauthorblockA{$^1$School of Systems \& Enterprises, Stevens Institute of Technology, Hoboken, NJ, USA}
\IEEEauthorblockA{$^2$University of Alabama at Birmingham, AL, USA}
%Email: \{akhanfor, anammouc, hghazzai, ye.yang, ymassoud\}@stevens.edu
}
}

\maketitle
\begin{abstract}
In this paper, we propose a machine learning process for clustering large-scale social Internet-of-things (SIoT) devices into several groups of related devices sharing strong relations. To this end, we generate undirected weighted graphs based on the historical dataset of IoT devices and their social relations. Using the adjacency matrices of these graphs and the IoT devices' features, we embed the graphs' nodes using a Graph Neural Network (GNN) to obtain numerical vector representations of the IoT devices. The vector representation does not only reflect the characteristics of the device but also its relations with its peers. The obtained node embeddings are then fed to a conventional unsupervised learning algorithm to determine the clusters accordingly. We showcase the obtained IoT groups using two well-known clustering algorithms, specifically the $K$-means and the density-based algorithm for discovering clusters (DBSCAN). Finally, we compare the performances of the proposed GNN-based clustering approach in terms of coverage and modularity to those of the deterministic Louvain community detection algorithm applied solely on the graphs created from the different relations. It is shown that the framework achieves promising preliminary results in clustering large-scale IoT systems.
\end{abstract}

\blfootnote{\hrule
\vspace{0.2cm} This paper is accepted for publication in 63rd IEEE International Midwest Symposium on Circuits and Systems (MWSCAS'20), Springfield, MA, USA, Aug. 2020. \newline \textcopyright~2020 IEEE. Personal use of this material is permitted. Permission from IEEE must be obtained for all other uses, in any current or future media, including reprinting/republishing this material for advertising or promotional purposes, creating new collective works, for resale or redistribution to servers or lists, or reuse of any copyrighted component of this work in other works.}%

\begin{IEEEkeywords}
Internet of Things (IoT), clustering, deep learning, graph neural networks.
\end{IEEEkeywords}
\vspace{-0.2cm}
\section{Introduction}
\vspace{-0.1cm}
% General IoT and Data anayltics
Internet-of-things (IoT) becomes essential in a variety of civil, public, and military applications, which makes their complexity and size perpetually increasing~\cite{al2015internet}. The growing number of connected devices requires advanced forms of collaboration to exploit their heterogeneity and improve their services effectively. The Social Internet-of-things (SIoT) concept has been emerged by allowing IoT devices to establish their own social networks~\cite{atzori2012social}. The paradigm aims to aid the smart objects to establish and maintain relations with their peers. The relationships in the network are not exclusive to machine-to-machine but can be extended between the users of the SIoT system, such as machine-to-human or even human-to-human relations. The social relations help assure trustworthiness between devices as the basis to share resources or collaborate on different services such as the share of computational needs. In fact, the relations between IoT devices may reflect their ownership, location, or past collaboration. However, understanding the structure of such complex and ubiquitous networks composed of diverse communicating nodes remains a challenging task. Novel data and graph analysis techniques can constitute an appealing solution to discern the SIoT network patterns and correlations among IoT devices~\cite{khanfor2019application}. 

Machine learning techniques can be used for this purpose to help in designing predictive analytics' techniques for SIoT networks and make well-informed decisions accordingly. For example, data analytics can help process, understand, and enhance the data generated by the devices~\cite{mahdavinejad2018machine}. It can also help understand the structure of IoT systems using unsupervised machine learning approaches such as classification and clustering methods to group IoT 1) infrastructures, e.g., by clustering devices to reduce the complexity of the vast IoT network or 2) services, e.g., by assigning IoT devices to tasks/services~\cite{mahdavinejad2018machine}. For example, the study of~\cite{liu2018identifying} employed machine learning to identify suspicious network activities by analyzing the transmission paths between the nodes.
%cite Najm paper if accepted
Another example is presented in~\cite{shao2018dynamic} where clustering algorithms are used as a first stage to reduce the complexity of a dynamic network of IoT devices.

\begin{figure}[t]
    \centering
    \includegraphics[width=\columnwidth]{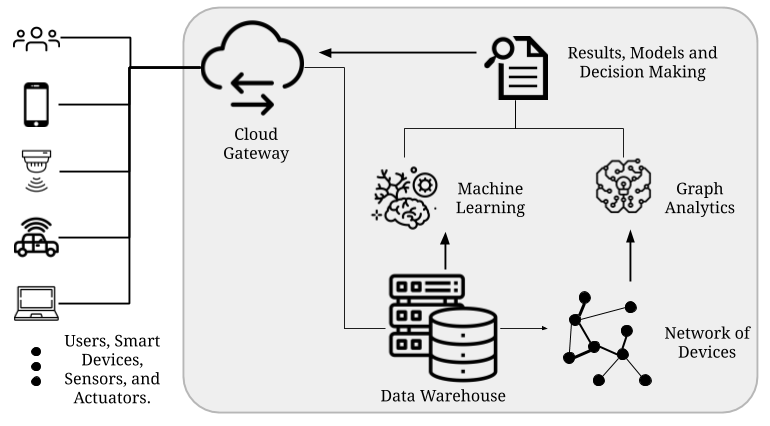}\vspace{-0.2cm}
    \caption{Data and network analytic architecture of IoT system.}
    \label{fig:ArchIoT}
    \vspace{-0.6cm}
\end{figure}

\begin{figure*}[!ht]
    \centering
    \includegraphics[width=\textwidth]{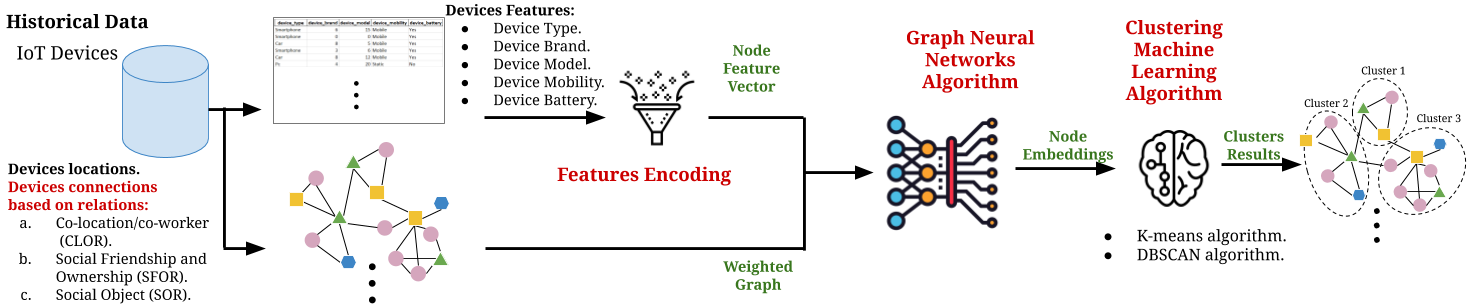}\vspace{-0.2cm}
    \caption{Graph Neural Network and clustering framework of SIoT devices.}
    \label{fig:framework}
    \vspace{-0.4cm}
\end{figure*}

In this paper, we develop a novel clustering approach based on Graph Neural Network (GNN), a deep learning algorithm, to discern SIoT structure. We aim to embed the features of devices as well as their connections from a real-world dataset using GNN and then apply an unsupervised learning algorithm to determine clusters of IoT devices sharing strong social relations. Results of GNN-based clustering are compared to the deterministic community detection approach, namely the Louvain method~\cite{khanfor2020automated}. In Fig.~\ref{fig:ArchIoT}, we illustrate a general SIoT data analytic framework where graph analysis and machine learning techniques are used to perceive the structure of SIoT system and the relations among its nodes. The IoT devices connect through a cloud gateway to exchange necessary data such as the location and specification of the devices. From this information and other IoT devices' features, graphs modeling the different social relations between the devices can be established. GNN and unsupervised learning techniques are then employed to determine the clusters of devices sharing strong social relations, which can help better understand the structure of the network and use this extra level of knowledge for more effective service discovery or mobile crowdsourcing~tasks.

\section{Proposed GNN-based Clustering Framework}
In Fig.~\ref{fig:framework}, we present the different steps of our proposed GNN-based clustering framework for social IoT systems. Starting from a dataset of $N$ IoT devices that includes several features such as Device Type, User ID owner, Device Brand, Device Mobility, Device Battery, and Device Geo-location, a pre-processing step is executed to create multiple weighted graphs of social relations connecting the devices. Afterwards, a GNN algorithm is applied to embed the nodes and their connections with numerical vector representations~\cite{kipf2016semi}. The GNN approach is enhanced such as it is capable of handling weighted graphs, i.e., the strength level of the social relations. Finally, an unsupervised clustering algorithm is applied on the vector presentations of the nodes to determine the different clusters of the IoT network.% apply the result of GNN two clustering algorithms $K$-means~\cite{louhichi2014density} and DBSCAN.
% based on number of features for both the requester device and the suitable cluster of capable devices.

\subsection{Social Relations and Data Pre-processing}
\subsubsection{Devices Relations}\label{relations}
There are different social relations between SIoT devices~\cite{atzori2012social}. These relations are based on the information about the devices such as ownership and geographical locations. In this study, we consider the following three social relations:

$\bullet$ \textit{Co-location/co-work based relation (CLOR):} This relation is inferred from the spatial features of the devices. Therefore, if there is a set of devices within a certain location, there are CLOR relations between these devices. The devices can be stationary or moving to different places. Therefore, mobile devices can dynamically change their CLOR links with other devices.

$\bullet$ \textit{Social object relation (SOR):} The SOR relation is created when two devices collaborate in a continuous or sporadic form. The criteria for setting the links are based on the owners' policies. For example, if two devices are co-located and exchange data for a certain period, then a SOR relationship can be established between them.

$\bullet$ \textit{Social friendship and ownership relation (SFOR):}
This relation is based on the social network of the owners and the ownership of the devices. Thus, we create high-weight links between devices that have common owners. The social network of owners can be then used to establish less weighted links among devices based on the number of friends to reach each owner (i.e., "friend" or "friend of a friend") and then project that on the SIoT network.

All the above relations in SIoT can be modeled by undirected and weighted graphs. The nodes are the devices of the IoT system and the edges are the social relations between these devices. The corresponding weights indicate the strength of social relations. The graphs do not include self-looped links on the objects.

\subsubsection{Features Encoding}
To ensure that the features of the devices in the dataset are suitable for the machine learner, we encode the categorical attributes such as Device Type, Brand, Mobility, and Battery using a one-hot encoder in Sci-kit pre-processing~\cite{scikit-learn}. The one-hot encoder transforms nominal data points to integer representation with consideration to limit the natural ordering comparing to the label encoding method. Moreover, the categorical textual values are encoded to integer values that will distinguish the data points, in which many clustering algorithms can handle. For example, in device type, there are a number of classes such as Smartphone, Smart Fitness, Pc, Car, etc. These types will be represented in integer values rather than a string. The resulting feature vector of each device $j$ of size $d\times 1$ is denoted by $\boldsymbol{X}_{j}$ where $j=1,\dots, N$ and $d$ is the number of features per node.

\subsection{Graph Neural Network Algorithm}\label{GNN_weight}
With the increase of computational power, many problems are represented by graphs. There is an emergence of adopting neural networks for graph classification, in general. The GNN surpasses that with the ability to handle a graph representation of nodes and edges to classify these nodes~\cite{scarselli2008graph}. This allows a better representation of the nodes and their relations by jointly embedding their features and their relationships with other IoT devices. The model follows a recursive neighborhood aggregation scheme, where each node aggregates feature vectors of its neighbors to compute its new feature vector. Thus, the node is represented by its transformed feature vector, which captures the structural information within its neighborhood and uses the nodes' different attributes as latent feature representations to enhance the learned representation. Given the weight matrix $\boldsymbol{A}$ of a social relation graph of $n$ nodes, we first normalize it to obtain the matrix $\boldsymbol{\tilde{A}}$ as follows:
\begin{equation}\label{first}
    \boldsymbol{\tilde{A}}= {\boldsymbol{\hat{D}}}^{-\frac{1}{2}}\boldsymbol{\hat{A}}{\boldsymbol{\hat{D}}}^{-\frac{1}{2}},
\end{equation}
where $\hat{\boldsymbol{A}} = \boldsymbol{A} + \boldsymbol{I}$ and $\hat{\boldsymbol{D}}$ is a diagonal matrix such that $\hat{D}_{ii}=\sum_{j}\hat{A}_{ij}$ and $\boldsymbol{I}$ is the diagonal matrix. The formula given in~\eqref{first} adds self-loops to the graph and normalizes each row of the emerging matrix $\boldsymbol{A}$. This normalization addresses numerical instabilities which may lead to exploding/vanishing gradients when used in a deep neural network model. Our GNN model consists of two message passing hidden layers, where the first hidden layer has $h_{1}^{}=64$ units and the second hidden layer has $h_{2}^{}=32$ units such as: 
\begin{equation}\label{first}
    \boldsymbol{Z}^{t}= f(\boldsymbol{\tilde{A}}\boldsymbol{Z}^{t-1}\boldsymbol{W}^{t}),
\end{equation}
where $\boldsymbol{\tilde{A}}$ is the normalized weight matrix of the graph given in (1), $\boldsymbol{W}^{t}$ is a matrix of trainable weights at layer $t$ such as $\boldsymbol{W_{}^{1}\in\mathbb{R}_{}^{d \times h_{1}}}$ and $\boldsymbol{ W_{}^{2}\in\mathbb{R}_{}^{h_{1} \times h_{2}}}$ , $f$ is the rectified linear activation function (ReLU), and $\boldsymbol{Z}^{t}$ is the learned embeddings of the graph in the $t^{th}$ layer. As an initialization, $\boldsymbol{Z}_{}^{0}=\boldsymbol{X}$ where $\boldsymbol{X\in\mathbb{R}_{}^{n \times d_{}}}$ is a matrix whose $j_{}^{th}$ row contains the feature vector $\boldsymbol{X}_{j}^{}$ of the IoT node $j$. The two message passing layers are followed by a fully-connected layer which makes use of the softmax function to produce a probability distribution over the class labels. 

Generally, the GNN model can be used as node classifier for either supervised or semi-supervised classification. But, given the unlabeled data in our case, we tend to use the GNN model as an embedder where we extract the feature representations of IoT devices in a forward pass using the propagation rule. We then label few nodes of our data and we train the model in a semi-supervised way in order to learn better representations.
In fact, the model is trained as a classifier yet we tend to only make use of the nodes’ hidden representations $\boldsymbol{Z_{}^{2}\in\mathbb{R}_{}^{n \times h_{2}}}$ that are produced from the second message passing layer. We use fixed hyper-parameters for all the graph sizes: learning rate equal to $10_{}^{-2}$ and dropout rate equal to $0.5$. We train the model over $100$ epochs.
Finally, we feed the extracted embeddings to unsupervised machine learning clustering algorithm to determine the communities and discover more clusters.
%In a forward pass we simply initialize the weights at random to produce the feature representations. We then can either semi-label the graph and train the GNN model to enhance the node representation learning or directly feed the embeddings to the clustering algorithm.
%Each node in the graph has a set of features $x_v$, and it ground-truth class $t_v$. The GNN is considered a semi-supervised algorithm by learning the class of the nodes partially and then classifying the unlabeled nodes. Each node will be represented in dimensional vector state $h_v$ that also contains the information of its neighbors that described in the following function~\ref{first}.
%\begin{equation}\label{first}
%    h_v = f(x_v,x_{co[v]},h_{ne[v]},x_{ne[v]})
%\end{equation}
%The $x_{co[v]}$ represents edges features. The $h_{ne[v]}$ denotes neighbours embedding of node $v$. $x_{ne[v]}$ denotes neighbours features of node $v$. And the function $f$ is the transition function represented in d-dimensional space. By implementing Banach fixed point theorem to $h_v$ and the previous equation~\ref{first} will be iteratively updated. This operation is also known as neighborhood aggregation or message passing.
%\begin{equation}\label{second}
%    \text{H}^{\textit{t}+1} = %F(\text{H}^{\textit{t}},\text{X})
%\end{equation}
%$\text{H}$ and $\text{X}$ is the concatenation of all $h$ and $x$. The result the algorithm is computed by giving the state $h_v$ and the element $x_v$ to get the result of $g$.
%\begin{equation}\label{three}
%    \text{O}_v = g(\text{h}_v,\text{x}_v)
%\end{equation}
\vspace{-0.1cm}
\subsection{Clustering Algorithms}
\vspace{-0.1cm}
Once a vector representation for each node in a social relation graph is determined using GNN, an unsupervised machine learning technique can be utilized to group the IoT devices with common features and attributes into clusters or communities. The similar IoT devices sharing strong social relations will be labeled in a cluster, while devices in different groups will have dissimilar features. In our study, we examine two clustering algorithms, namely the $K$-means and Density-Based Spatial Clustering of Applications with Noise (DBSCAN) algorithms.

\subsubsection{$K$-means} \label{kmeans}
It is one of the wide used unsupervised clustering algorithms~\cite{arthur2006k}. It achieves the clustering by pre-specifying the number of clusters, $K$. In general, $K$-means iterate to determine $K$ virtual centroids to which it associates the closest data points using the sum of the squared distance separating the vectors. It converges when no further enhancement is achieved. The $K$-means algorithm aims to choose centroids that minimize the inertia or sum-of-squares within-cluster criterion.
\subsubsection{DBSCAN} \label{dbscan}
The main objective of DBSCAN is to identify the clusters based on the density of data points~\cite{ester1996density}, where the most similar data points are dense together and form a distinguished group where the different clusters are separated with less density data points. The algorithm starts with an arbitrary point and converges when all the data points are visited. It uses a distance threshold to decide whether a nearby point belongs to the same cluster or not. If not, it will assigned as a noise, which can be part to another cluster. The previous process will be repeated over all the data points until the density-connected cluster is achieved.

Once the clustering algorithm is run for each SIoT relation, groups of devices sharing strong social relations are determined. The IoT devices may then cooperate together in a trustworthy manner.

\section{Results \& Discussions}\label{sec:results}
To examine the proposed framework, we use a data set of real-world IoT devices from a smart city in Santander, Spain, provided by Marche et al.~\cite{marche2018dataset}. The data set includes different types of private and public devices. We select 1000, 1500, and 2000 private devices out of 16216 devices to analyze the possibility of the framework for scalability and applicability in different sizes of the IoT system. Following that, the links between the devices are established based on the various social relations, namely CLOR, SFOR, and SOR, described in Section~\ref{relations}.

To assess the quality of the different clustering results, we use two of the standard cluster quality metrics in our study: modularity and coverage. Graph modularity analyzes the presence of each intra-cluster edge of the graph with the probability that that edge would exist in a random graph. It is expressed as follows: %~\cite{4358966}
\begin{equation}\label{first}
    \text{Q} = \frac{1}{2m}\sum_{vw}^{}\left({A_{vw}^{}}-{\frac{k_{v}k_{w}}{2m}}\right)\delta(c_{v}c_{w}),
\end{equation} 
where $\delta$ is the Kronecker delta, it equals to one if u and v belong to the same community and 0 otherwise, $k_{u}$ is the degree of node $u$, $m$ is the number of edges in the graph, and $A_{vw}$ is the element located at row $v$ and column $w$ of the adjacency matrix $\boldsymbol A$.

%\begin{equation}\label{first}
%    \text{Mod} = %\sum_{k}^{}({e_{kk}^{}}-{a_{k}^{2}}),
%\end{equation}
%where ${e_{kk}^{}}$ is the probability of intra-cluster edges in cluster ${S_{k}^{}}$, and ${a_{k}^{}}$ is the probability of either an intra-cluster edge in cluster ${S_{k}^{}}$ or of an inter-cluster edge incident on cluster ${S_{k}^{}}$. They are expressed as follows:
%\begin{equation}\label{first}
%     {e_{kk}^{}}= |\{(i,j) : i\in S_{k}^{}, j\in S_{k}^{},(i,j)\in E\}|/|E|
%\end{equation}
%\begin{equation}\label{first}
%     {a_{k}^{}}= |\{(i,j) : i\in S_{k}^{} ,(i,j)\in E\}|/|E|
%\end{equation}
%where $E$ is the set of edges of the graph.\\
As for coverage metric, it compares the fraction of intra-cluster edges in the graph to the total number of edges in the graph. It is given by:%~\cite{eurovisshort.20141153}
\begin{equation}\label{first}
     \text{Cov} = \frac{\sum_{i,j}^{} A_{ij}^{}\delta(S_{i}^{}, S_{j}^{})}{\sum_{i,j}^{}A_{ij}^{}},
\end{equation}
where ${S_{i}^{}}$ is the cluster to which node $i$ is assigned and $\delta(a, b)$ is 1 if $a = b$, otherwise is equal to 0. Coverage falls in the range 0 to 1, and 1 is the highest score that indicates that a graph topology is well-clustered.

\begin{figure}[t]
    \centering
    \includegraphics[width=\columnwidth]{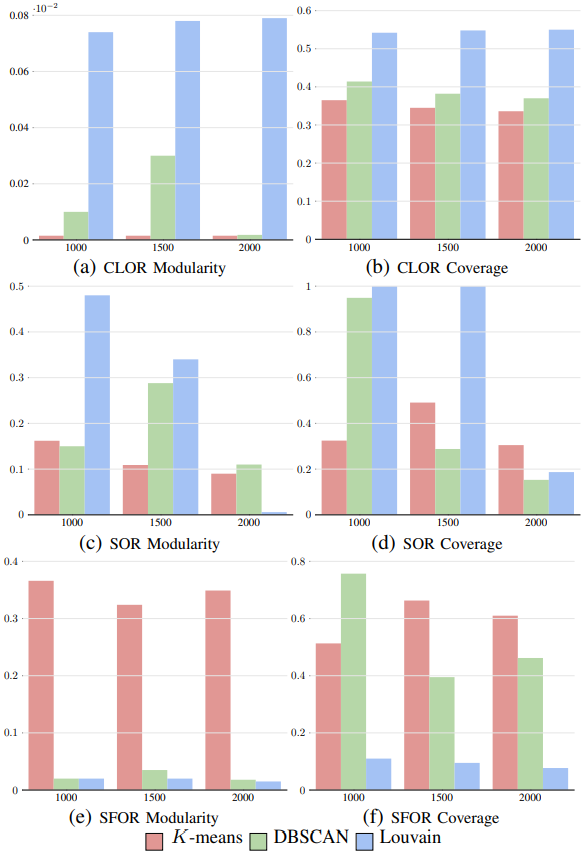}\vspace{-0.2cm}
    \caption{Modularity score and coverage presented for the $K$-means, DBSCAN, and Louvain community detection algorithms with three different social network connectivity scales (1000, 1500, and 2000).}
    \label{fig:ArchIoT}
    \vspace{-0.6cm}
\end{figure}

%We use the GNN model as a node embedder to get the node representations. At first, in a forward pass, we initialize the weights at random to produce the feature representations. We then can either semi-label the graph and train the GNN model to enhance the node representation learning or directly feed the embeddings to the clustering algorithms. We notice the ability of the GNN model to perform informative node embeddings even with randomly-initialized weights. After that, We used an API from Scikit learn for unsupervised clustering algorihtms\cite{scikit-learn} to implement $K$-means and DBSCAN on our study. 
In Fig. 3, we illustrate the clustering results of applying algorithms $K$-means, DBSCAN, and Louvain for the evaluation metrics, modularity, and coverage. We compare the obtained results to one of the deterministic Louvain algorithms. Each sub-figure presents one of three graphs; CLOR, SOR, and SFOR, with three different IoT networks, scale 1000, 1500, and 2000 nodes. The $K$-means is executed using the elbow method to determine the best number of clusters. However, we notice that $K$-means present lower performance when directly applied to the node embeddings of the GNN model, shown in Fig. 3 as red bars. In fact, $K$-means aims to choose centroids that minimize the inertia, which is not a normalized metric. We know that lower values are better, and zero is optimal. But, in very high-dimensional spaces, Euclidean distances tend to become inflated (this is an instance of the so-called “curse of dimensionality”). We run a dimensionality reduction algorithm, i.e., t-distributed Stochastic Neighbor Embedding (T-SNE), before $K$-means clustering, which alleviates this problem, speeds up the computations and presents a visualization method in the 2-dimensional space for the clusters.

With DBSCAN, we notice that some nodes are detected as outliers. Therefore, we assign to each of those nodes a new cluster. We observe that the $K$-means gives well-separated clusters but tends to restrict the number of the groups comparing to DBSCAN. We also notice that both K-means and DBSCAN clustering of the GNN embeddings outperform Louvain community detection mainly for the SFOR network in all scales and the SOR network in large size (2000 nodes). Despite its performance with the CLOR network comparing to the two other methods, the Louvain algorithm tends to restrict the discovered communities to two clusters for the three different scales.

In Table~\ref{tab:results_clusters}, where a comparison based on the numbers of clusters obtained for each relation with different networks sizes. The number of clusters for DBSCAN tends to be higher than the other methods for all the networks. This characteristic remains the same even when we do not consider the outliers as separated clusters. Finally, the whole process from the embedding and clustering to the dimension reduction is relatively fast compared to the Louvain method, which is an advantage when applying our approach to a vast network of devices.
%In Table~\ref{tab:results_clusters}, the number of clusters tend to be higher than the other methods caused by detecting a number of nodes as outliers. The green bars in Fig.~\ref{fig:ClusteringPerformance} indicate that DBSCAN underperforms one of the other clustering algorithms we used in this analysis.

\begin{table}[t]
\caption{Obtained number of clusters}
  \centering
  \begin{adjustbox}{max width=\columnwidth}
\addtolength{\tabcolsep}{-5pt}\begin{tabular}{c|ccc|ccc|ccc}
\toprule
            & \multicolumn{3}{c}{\textbf{CLOR}}                   & \multicolumn{3}{c}{\textbf{SFOR}}   & \multicolumn{3}{c}{\textbf{SOR}}                 \\
No. Devices & $K$-means       & DBSCAN      & Louvain      & $K$-means     & DBSCAN       & Louvain  & $K$-means     & DBSCAN       & Louvain    \\
\midrule
1000        &     4         &     19  &      2   &      5       &    47   &  9   & 5 & 50 & 18      \\
1500        &     7         &    105  &      2   &      11      &    181  &  12  & 7 & 73 & 18      \\
2000        &     8         &    551  &      2   &      15      &    247  &  14  & 10 & 228& 25      \\ %\midrule
 %           & \multicolumn{6}{c}{\textbf{SOR}}                                                                \\
  %          & \multicolumn{2}{c}{$K$-means} & \multicolumn{2}{c}{DBSCAN} & \multicolumn{2}{c}{Louvain} \\ \midrule
%1500        & \multicolumn{2}{c}{7}        & \multicolumn{2}{c}{73}       & \multicolumn{2}{c}{18}        \\
%2000        & \multicolumn{2}{c}{10}        & \multicolumn{2}{c}{228}       & \multicolumn{2}{c}{25}      
\bottomrule
\end{tabular}
\end{adjustbox}\vspace{-0.4cm}
\label{tab:results_clusters}
\end{table}
\vspace{-0.1cm}
\section{Conclusion}
In this study, we propose a novel GNN-based clustering approach for SIoT devices having different social relations. For a real-world dataset, we embed the features of the nodes as well as their social relations using GNN and then, feed the obtained vector representations to a conventional clustering algorithm to determine communities of socially connected IoT devices. The process allows fast conversions of complex IoT systems into structured groups of devices that can be exploited to enhance the discovery and object identification for various IoT applications. We notice that different clustering algorithms, case by case, can outperform other community detection methods for certain metrics such as modularity and coverage, which represents a promising result to further examine several machine learners. As future work, we will focus on online learning approaches for community detection to consider mobile and dynamic IoT networks.
\vspace{-0.1cm}
\bibliography{references}
\bibliographystyle{ieeetr}
\balance

\end{document}